\documentstyle[12pt,epsfig]{article}

\begin{document}
\title{Rotating Black Hole, Twistor-String and Spinning Particle}
\author{{Alexander Burinskii }\\
\\
{\it NSI Russian Academy of Sciences,} \\
{\it B.Tulskaya 52, Moscow 115191, RUSSIA} \thanks{Talk given at the
Conference `Symmetries and Spin'(SPIN-Praha-2004) July 2004.}}


\maketitle

\def\b{\bar}
\def\d{\partial}
\def\D{\Delta}
\def\cA{{\cal A}}
\def\cD{{\cal D}}
\def\cK{{\cal K}}
\def\f{\varphi}
\def\g{\gamma}
\def\G{\Gamma}
\def\l{\lambda}
\def\L{\Lambda}
\def\M{{\cal M}}
\def\m{\mu}
\def\n{\nu}
\def\p{\psi}
\def\q{\b q}
\def\r{\rho}
\def\t{\tau}
\def\x{\phi}
\def\X{\~\xi}
\def\~{\tilde}
\def\h{\eta}
\def\bZ{\bar Z}
\def\cY{\bar Y}
\def\bY3{\bar Y_{,3}}
\def\Y3{Y_{,3}}
\def\z{\zeta}
\def\Z{{\b\zeta}}
\def\Y{{\bar Y}}
\def\cZ{{\bar Z}}
\def\`{\dot}
\def\be{\begin{equation}}
\def\ee{\end{equation}}
\def\bea{\begin{eqnarray}}
\def\eea{\end{eqnarray}}
\def\half{\frac{1}{2}}
\def\fn{\footnote}
\def\bh{black hole \ }
\def\cL{{\cal L}}
\def\cH{{\cal H}}
\def\cF{{\cal F}}
\def\cP{{\cal P}}
\def\cM{{\cal M}}
\def\ol{\overline}
\def\const{{\rm const.\ }}
\def\ik{ik}
\def\mn{{\mu\nu}}
\def\a{\alpha}

\begin{abstract}

We discuss basic features of the model of spinning particle based on
the Kerr solution. It contains a very nontrivial {\it real} stringy
structure  consisting of the Kerr circular string and an axial
stringy system.

We consider also the complex and twistorial structures of the Kerr
geometry and show that there is a {\it complex} twistor-string
built of the complex N=2 chiral  string with a twistorial
$(x,\theta)$ structure. By imbedding into the real Minkowski $\bf M^4$,
the N=2 supersymmetry is partially broken and string
acquires the open ends. Orientifolding this string, we identify
the chiral and antichiral structures. Target space of this
string is equivalent to the Witten's `diagonal' of the $\bf
CP^3\times CP^{*3}.$
\end{abstract}

\section*{Introduction}

In the recent paper \cite{WitTwi} Witten suggested a new fundamental
object, twistor-string, incorporating twistors in superstring
theory. It was conjectured that the twistor-string controls the
scattering processes and is responsible for analytic properties of
the scattering amplitudes, and therefore, this string
may be an element of the structure of fundamental particles. This
Witten paper and the subsequent papers have paid extraordinary
attention which has been heated by the progress in computation of some
scattering amplitudes on the base of twistor inspired methods
\cite{Nai,ChaSie}. In particular, the striking simplifications of the
maximally helicity violating (MHV) amplitudes were achieved by the
calculations performed on the base of the spinor helicity formalism
(for review see for example \cite{WitTwi,BDK}).

In this paper we consider the model of spinning particle
\cite{BurTwi} which is based on the Kerr rotating black hole
solution. The Kerr geometry has a remarkable twistorial structure,
and the Kerr's spinning particle has also a very non-trivial stringy
structure. The aim of this paper is to describe the complex,
twistorial and stringy structures of the Kerr spinning particle and
to argue that the new Witten's  `twistor-string' is related to
the structure of the Kerr spinning particle, and therefore,
the fusion of the twistor and string theories  has a natural extension
to the structure of the rotating black holes.

In Sec.2 we give a very brief introduction to twistors, in.Sec.3 we
treat briefly the structure of the Kerr spinning particle and its
basic properties. In sec.4 the twistorial structure of the Kerr
geometry is discussed. In Sec.5 we consider the complex Kerr
geometry and show that it contains a version of the complex
twistor-string which has some properties of the well known N=2
string \cite{OogVaf} and also the chiral $(x,\theta)$ twistor-string
\cite{Ber,Sie24,LecPop,Nai2}. Being imbedded into $\bf M^4$, this
string turns out to be open with a partially broken N=2
supersymmetry. The world-sheet of this string is formed by
orientifolding the chiral and antichiral
structures, and target space turns out to be
equivalent to Witten's `diagonal' of the $\bf CP^3 \times CP^{*3}$.

\section{Twistors}
Twistor may be considered as a world-line of the lightlike particle,
which geometrically is a null line in Minkowski spacetime ${\bf
M^4}$ \cite{Pen}.
Momentum of the lightlike particle is described by the null
vector $p^\m$ satisfying the condition $p^\m p_\m=0$ and may be
represented in the spinor form
\be p^\m = \psi \sigma ^\m \bar \psi
\ , \label{psi}
\ee
where $\sigma^\m$  are the Pauli matrices. A
null line going via the coordinate origin may be described as $x^\m
(t)= p^\m t$.

We will also consider the complexified Minkowski space ${\bf CM} ^4 $.
In this case
$p^\m = \psi \sigma ^\m \tilde \psi $, where $\tilde \psi$ and $\psi $
are independent spinors.  Varying $\tilde \psi $ at fixed $\psi$  one
obtains the (say ``left'' ) complex planes.
Similar, varying $\psi$ at fixed
$\tilde \psi $ one obtains another family of the (``right'') complex planes.
They are totally null planes in the sense that any vector lying in them
is null. Therefore,
from the complex point of view twistors are complex null planes.
Intersection of the complex conjugated
left and right planes yields the real twistor.

{\it The main difference between twistors and spinors} is that
twistors describe the null lines which has a definite position
with respect to the origin of coordinates (or an observer).
Correspondingly, the extra
parameters are necessary to fix the position of line.
If a point $ x_0^\m $ is fixed on the line,
$x^\m (t) =  x_0^\m +  p^\m t$, it may be described by two  spinors: $\psi
^\alpha$ and $\omega _{\dot \alpha}= x_0^\n \sigma _{\n \alpha
\dot \alpha}$. In the flat spacetime these parameters  are
independent of the position $x_0^\n$ on the null line, and
the four spinor parameters
\be Z^a =\{ \psi ^\alpha, \omega _{\dot \alpha} \} \ee
give the basic representation of twistor.
However, the initial set
\be \{ x_0^\m, \ \psi ^\alpha \} \label{xpsi} \ee
may also be considered as an equivalent representation which is
used in the {\it spinor helicity formalism} \cite{BDK}.
Finally, since spinor $\psi ^\alpha $ gives the homogenous
coordinates of the null ray, one can use the equivalent
projective spinor coordinate  $Y = \psi _2 / \psi _1  \in {\bf CP^1} $
, and represent twistor in the form
\be \{ x_0^\m, \ Y \},
\label{xY}
\ee
or via tree projective twistor coordinates
\be
 \{Y=Z^2/Z^1, \quad \l _1 = Z^3/Z^1, \quad \l _2 =Z^4/Z^1\} \in {\bf CP^3} .
\label{PTw}
\ee
The both forms  (\ref{xY}) and  (\ref{PTw}) are used in the Kerr-Schild
formalism \cite{DKS} and in the
description of twistor-string \cite{WitTwi}.

\section{The Kerr spinning particle: black hole which is neither
`black' nor `hole'}

The joke in the title of this section belongs to P. Townsend. It
has direct relation to the Kerr spinning particle.

The Kerr geometry  has  found application in a very wide range of
physical systems: from the rotating black holes and galactic nucleus
to  fundamental solutions of the low energy string theory. It must
not be wonder that it has the relations to spinning particles
too. Indeed, the Kerr-Newman solution has anomalous, $g=2$,
gyromagnetic ratio of the Dirac electron.
In the units $\hbar =c=G=1, $ the parameters of
the Kerr solution adapted to electron are $e^2 \sim 1/137, \quad
m\sim 10^{-22}, \quad a\sim 10^{22}, \quad ma=1/2.$

\begin{figure}[ht]
\centerline{\epsfig{figure=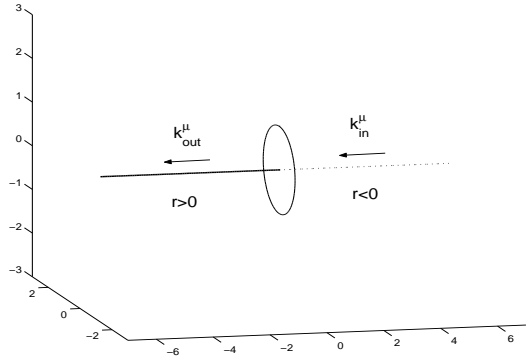,height=5cm,width=7cm}}
\caption{The Kerr singular ring and a semistring (or semitwistor)
emerging from the `negative' sheet of the Kerr space.}
\end{figure}

One sees that $a>>m$, and the Kerr black-hole horizons disappear, revealing
the naked singular ring of the Compton radius $ a\sim \hbar/m .$
Metric is almost flat and gravitational field is concentrated in a very
narrow vicinity of the singular ring.

  The Kerr singular ring is the branch line of the
Kerr space on two sheets, `positive'  ($r>0$) and `negative'
($r<0$), where $r$ is radial coordinate
of the oblate spheroidal coordinate system.
The fields change directions and signs on the negative sheet.
The negative sheet acquires an interpretation of the
sheet of advanced fields which may be related to the vacuum
zero point field \cite{BurOri,BurNst,BurAxi}.
The particles, strings or twistors disappear in (emerge from) a mirror
world  passing through the Kerr ring (see Fig. 1).

In the old paper \cite{Bur0} a model of the spinning Wheeler's geon
was suggested, in which a quantum electromagnetic excitation -
traveling waves along the Kerr ring generate the spin and mass of
the geon. It was recognized soon \cite{IvBur} that the Kerr ring
is a closed chiral string.

The treatment of singular lines as strings is very natural. Among the well
known examples are the fundamental strings of the low energy string theory
\cite{DGHR,Sen} and the strings
interacting with Higgs field, in particular, the Nielsen-Olesen vortex
line in a superconducting media and  the Witten chiral superconducting
cosmic string, where one of two Higgs fields is concentrated on the singular
line \cite{Wit}.
It was shown in \cite{BurSen} that the field around the Kerr ring is similar
to the field around the fundamental heterotic string, and finally, in
\cite{BurOri}  that the Kerr ring is a D-string with an orientifold
world-sheet.

Recently, we have shown \cite{BurTwi,BurAxi} that the circular Kerr string
is not unique in the
Kerr spinning particle, and there is a second, `axial' stringy system.
It was shown that the wave excitations  of the Kerr ring lead unavoidable
to the
appearance of the extra axial singular lines which are chiral strings
similar to the Witten cosmic strings.
Therefore, the Kerr spinning particle
acquires a stringy skeleton forming by the Kerr circular string and by
the axial stringy system consisting of two semi-infinite strings of opposite
chiralities, see Fig. 2.

The axial semistrings are not new objects in the Kerr geometry.
In fact, the NUT-parameter of the Kerr-NUT solution led to the appearance of
a semistring carrying the magnetic flow. Moreover, the general class of the
Kerr-Schild solutions \cite{DKS}
contains a free holomorphic function $\psi (Y)$ which defines a series of
the {\it exact} stationary solutions of the Einstein-Maxwell system
containing singular semistrings.

In the simplest cases $\psi(Y) \sim Y^n $, where $n$ is integer\fn{
Half-integer index $n$ can be considered in supergravity and also by
twisting of spin-structure, see note in the Conclusion.}, and $Y$ has
the form
\be Y = e^{i\phi} \tan \frac \theta 2
\label{Yspher}
\ee
in the Kerr oblate spheroidal coordinates.
By $n <0$ solutions are singular at $\theta =0$, along the positive
semiaxis $z^+$, and by $n >0$ solutions are singular at $\theta =\pi$,
which corresponds to negative semiaxis $z^-$.
The case $n=0$ corresponds to the charged Kerr-Newman field, and the
axial singularity is absent.
Up to our knowledge these solutions never were analyzed.
In the case of spinning particle we have solutions with similar topological
structure of singularities, however the
function $\psi$ acquires an extra
dependence on the retarded-time parameter $\t$,

$\psi \sim Y^n e^{i\omega \t}.$

As a result
there appear traveling waves along the Kerr circular string and the chiral
traveling waves along the axial semistrings\cite{BurAxi,BurTwi}.

It was shown that the chiral excitations of the axial stringy system may be
described by the Dirac equation. In the Weyl basis the Dirac current is
represented as a sum of two lightlike components of opposite
chiralities

$ J_\m = e (\bar \Psi \gamma _\m \Psi) = e (\chi ^{+} \sigma _\m  \chi +
\phi ^{+} \bar \sigma ^\m  \phi ).$

The four-component spinor
$ \Psi =
\left(\begin{array}{c}
\phi _\alpha \\
\chi ^{\dot \alpha}
\end{array} \right),$ satisfies the massive Dirac equation which describes
an interplay
of two traveling waves of opposite chiralities - the lightlike fermions
trapped by two axial singular semistrings. De Broglie periodicity appears
as a result of the beating of these waves.

This axial stringy system plays an important
role in the formation of the Kerr's twistor-string. We will show that
two axial semistrings play the role of two D1-branes which are boundaries
of the complex twistor-string.

It reproduces a very specific model of quarks: the fermionic
currents of opposite chiralities adjoined to the ends of the
twistor-string.

\bigskip

\begin{figure}[ht]
\centerline{\epsfig{figure=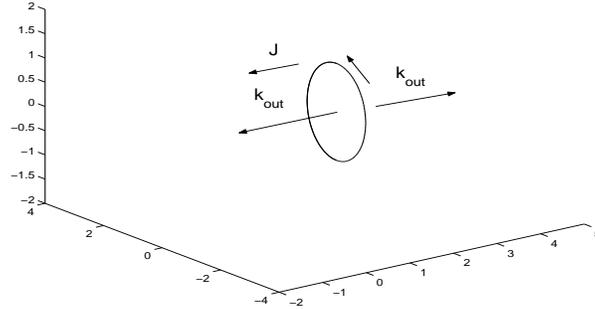,height=5cm,width=8cm}}
\caption{Stringy skeleton of the spinning particle. Circular string
and two semi-infinite D-strings of opposite chiralities.}
\end{figure}

{\bf We summarize bellow the basic features and properties of the Kerr
spinning particle:}
\begin{itemize}
\item Anomalous gyromagnetic ratio $g=2$ of the Dirac electron,
\item The mass $m$, spin $J$ and charge $e$ are the only free parameters,
\item Model is semiclassical - the quantum constant is used {\it only} to
quantize energy of excitations,
\item Particle has the Compton size $a=\hbar /m$, and the Compton region
 is stringy structured,
\item Nontrivial real stringy structure: topological coupling of the axial
and circular strings,
\item Mass and spin have the origin from the circular traveling waves,
electromagnetic excitations of the Kerr singular ring (the Wheeler's `geon'),
\item Axial stringy system is
carrier of  de Broglie waves which are described by the Dirac
equation,
\item Axial stringy system is responsible for the scattering at high energies,
\item Remarkable complex and twistorial structures, twistor-string.
\end{itemize}

The following properties may also have a relation to the foundations
of quantum theory\fn{They were discussed in previous papers, see
\cite{BurTwi,BurAxi} and reference therein}:

\begin{itemize}
\item  Wave function acquires a physical carrier - the axial stringy
system,
\item   Axial string  controls the motion of
particle due to topological coupling of the axial and circular
singular strings, so the model reproduces the features of the old
wave-pilot conjecture by de Broglie.
\item  The quantum property - absence of radiation by oscillations - is
exhibited here at the classical level, since the loss of energy by
stringy excitations is compensated by the ingoing radiation from
the `negative' sheet of the Kerr spacetime.
\end{itemize}

\section{Kerr geometry in the Kerr-Schild formalism}

\subsection{Real Kerr geometry and twistors}
In the Kerr-Schild approach  \cite{DKS},
the Kerr-Schild ansatz for metric is used
\be g_{\m\n}
= \h_{\m\n} + 2 h k_{\m} k_{\n}, \label{ksa} \ee
where $ \h_\mn $
is metric of auxiliary Minkowski space-time, $x^\m =(t,x.y.z)$.
\fn{We use signature $-+++$.}
Function
\be h= \frac
{mr-e^2/2} {r^2 + a^2 \cos^2 \theta},\ee
is written in the oblate spheroidal coordinates, and
$k_\m$ is a twisting
null field which is tangent to the Kerr principal null congruence
(PNC) which represents a family of the geodesic null rays - twistors -
covering twice the space-time.

Since $k^\m$ is null, it can be represented in the spinor form (\ref{psi}),
however, in the Kerr-Schild formalism the projective spinor coordinate $Y$
is used, and the Kerr twistorial congruence is determined by a complex scalar
function $Y(x)$, in terms of which the vector $k^\m$ takes the form
\be k_\m dx^\m = P^{-1}(du + \bar Y d \zeta + Y d \bar\zeta - Y \bar Y dv) ,
\label{kmu}
\ee
where $P=2^{-1/2}(1+ Y \bar Y)$ is a projective factor and
\bea
2^{1\over2}\z = x+iy , \quad 2^{1\over2} \Z &=& x-iy , \nonumber \\
2^{1\over2}u = z - t , \quad 2^{1\over2}v &=& z + t  \label{ncc}\eea
are the null Cartesian coordinates.

Therefore, for each point of the Kerr spacetime $x$ we have the pair
$\{x,Y(x)\}$ which determines a twistor in the form (\ref{xY}).

\begin{figure}[ht]
\centerline{\epsfig{figure=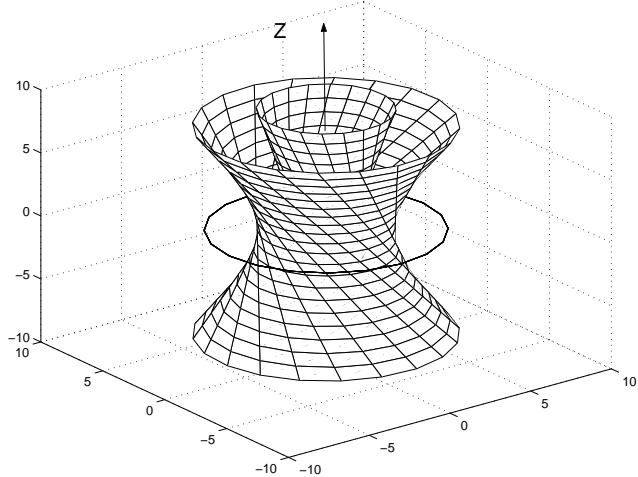,height=6.5cm,width=8.5cm}}
\caption{The Kerr singular ring and the Kerr
congruence of twistors (PNC). Singular ring is a branch line of
space, and PNC propagates from the ``negative'' sheet of the Kerr
space to the ``positive '' one, covering the space-time twice. }
\end{figure}
The Kerr's twistorial congruence is geodesic and shear-free.
Congruences of this type are determined  by {\it the Kerr theorem},
via the solution $Y(x)$ of the algebraic equation $F=0$, where
$F(Y,\lambda _1, \lambda _2)$ is an arbitrary holomorphic function
of the projective twistor coordinates \fn{The twistor form
(\ref{PTw}) is used here. Notice, that the Kerr theorem, which holds
one of the prominent positions in twistor theory \cite{Pen}, was
practically used by derivation of the Kerr-Schild class of solutions
\cite{DKS}, but it was never formulated by Kerr in the twistor terms
as well as in the form of a theorem. Meanwhile, the close relation
of twistors to the Kerr geometry is not incidental since these ideas
were discussed by Roy Kerr and Roger Penrose well before the
corresponding publications\cite{Ker}.}

$\{Y,\quad \lambda
_1 = \zeta-Y v,\quad \lambda _2=u+Y\bar\zeta\} = Z^a/Z^1 $.

Function $Y(x)$ allows one to reconstruct from (\ref{kmu}) the null vector
$k^\m$ which is tangent to the Kerr congruence. This vector is the principal
null vector which determine the metric, the Kerr-Newman vector field, and
the direction of radiation for radiative Kerr-Schild solutions.

One of the real null tetrad vectors of
the Kerr-Schild tetrad is taken proportional to $k^\m$, $e^3 =P k^\m$,
and tetrad is  completed as follows
\begin{eqnarray}
e^1 &=& d \zeta - Y dv, \qquad  e^2 = d \bar\zeta -  \bar Y dv, \nonumber \\
e^3 &=&du + \bar Y d \zeta + Y d \bar\zeta - Y \bar Y dv, \nonumber\\
e^4 &=&dv + h e^3\label{KSt},
\end{eqnarray}
where $e^1$ and $e^2$ are complex conjugate.

The caustics of the Kerr PNC (singular lines) are determined by the system
of equations $F=0, \ dF/dY =0$.
In the Kerr solution function $F$ is quadratic in $Y$, and these equation
can be explicitly
resolved \cite{BurNst} leading to the
function $ Y(x)$ in the form (\ref{Yspher}) which yields the
structure of the Kerr PNC and the Kerr
singular ring shown in Fig. 3.

The twistor $\{ x,Y \}$ is a complex null plane at
the point $x$ which is spanned by the null vectors $e^1$ and $e^3$.
The real null rays of the Kerr congruence are the real slices of these
null planes. The congruence is determined by the real field $k^\m(x)$
(or $e^{3\m}=P k^\m$) which describes a flow of
the lightlike field propagating from the `negative' sheet of the Kerr space
onto `positive' one through the disk spanned by the Kerr ring.
In particular, the Kerr PNC determines a flow of radiation along $e^3$, so
the outgoing radiation is compensated by an ingoing flow on the
negative sheet.

\subsection{Complex Kerr geometry and twistors}

The Kerr twistor-string appears naturally in
the Newman-initiated  {\it complex representation of the Kerr geometry}
\cite{New,BurStr,BurNst}, in which the Kerr solution is
generated by a complex source propagating along a complex world line
 $X_0^\m (\t) \in {\bf CM^4}.$
The Kerr-Schild formalism is well adapted to this representation,
since the Kerr-Schild ansatz (\ref{ksa}) assumes the auxiliary Minkowski
spacetime with Cartesian coordinates $x^\m$,
which may be complexified, forming coordinates of ${\bf CM^4}.$
The time parameter $\t$ is also complex in this representation,
$\t =t +i\sigma$.

The real fields on the real space-time $x^\m \in {\bf M^4}$ are determined
via a complex retarded-time construction, in which the key role is played
by the complex light cones, adjoined to the real points of observation
$x^\m$ and to the points of emanation on the complex world line $X_0^\m(\t)$.

It is well known that the real light cone may be split on two family of
null planes. There are different ways of doing this \cite{BurStr,BurNst},
and all the ways turns out to be close related to twistors.
The light cone at a point $x_0$ may be written in the
spinor form
\be {\cK}_{x_0}=\{x:  x^\m  = x_0^\m + \psi ^\alpha
\sigma ^\m _{ \alpha \dot \alpha} \tilde \psi ^\alpha \} \label{lKpsi} .\ee
The real null rays of the ${\cK}_{x_0}$ correspond to the complex conjugate
spinors
$ \psi ^\alpha $ and $ \tilde \psi ^{\dot\alpha}. $ As we discussed in sec.2,
varying $\tilde\psi $ at fixed $\psi$  one
obtains the (`left' ) complex planes, and
 varying $\psi$ at fixed $\tilde \psi $ one obtains
another family which we call the right complex planes.
We can also use the projective spinor coordinates $Y$ and $\bar Y$ and
describe the complex null cone as ${\cK}_{x_0}=\{x_0, \ Y, \ \bar Y \}$  and
 the left (right) null planes of the cone by twistor $\{x_0, \ Y \}$
(or $\{x_0, \ \bar Y \}$).
Notice, that the apex of the light cone $x_0$ may also be complex.

To form the retarded-time construction we
 set the apex of cone $x_0$ at a real point $x \in {\bf M^4}$ and
have to find the points of
intersection of this light cone with the complex world-line $X_0(\t)$ to
determine the value of the retarded-time parameter $\t _0$.
In the stationary Kerr case this world-line has the simplest form
of the straight timelike line $X_0^\m(\t)=(\t,0,0,ia).$
Nevertheless, at this point we meet a peculiarity which complicates the
usual retarded-time scheme: there may be four roots for the points of
intersection. We can ignore one pair of the roots assigning them to
the advanced fields. The residual two roots are linked with the fact
 that the left and right null planes yield the different points of
intersection with a complex world-line. Therefore, we must mark the roots
as the left retarded time $\t_L$ and the right one $\t_R$.
To get a definiteness we have to use for the roots (for example)
only the left null planes, i.e. the holomorphic twistor pairs $\{x, \ Y\}$.
It shows that twistors appear very natural in the complex retarded-time
construction.

\section{Complex Kerr string}

In accordance with the above complex representation of the Kerr geometry,
the Kerr solution is
generated by a complex source propagating along a complex world line
$X_0^\m (\t)\in {\bf CM^4}$.
Complex world-line is parametrized by complex time $\t=t+i\sigma$ containing
 two parameters $t$ and $\sigma$, and therefore, it describes a
world sheet. The complex world line  $X_0^\m (\t)$ can be considered as an
$N=2$ string in ${\bf CM^4}$\cite{OogVaf,BurStr,BurOri}.

The Lagrangian for this string is \cite{OogVaf,GibPer}
\fn{We omit the fermi terms.}
\be {\cL}=\int d\t d \bar\t (\eta _{i\bar j} ( \d _{\t} \bar X_0^i
\d _{\bar \t}  X_0^j + \d _{\bar \t} \bar X_0^i \d _{\t}  X_0^j ).
\label{hL}
\ee
As usual, the general solution is given as a sum of the left and right
modes, \be X_0^i (t,\sigma) = X_L^i (\t) + X_R ^i (\bar\t) ,\ee which are not
necessarily complex conjugates to each other.
The constraints take the form
\be \eta _{i\bar j} \d _{\t} \bar X_L^i \d _{\t}  X_R^j =0 ; \quad
\eta _{i\bar j} \d _{\bar\t} \bar X_L^i \d _{\bar\t}  X_R^j =0. \ee
Notice, that the mode expansion contains the hyperbolic basis functions which
are not orthogonal over the string length, so similar to the $N=2$ string
there are no excitations indeed. Meanwhile, one can see that the straight
complex world-line which describes the complex Kerr source,
\be X_0\equiv X_L^\m(\t)=X_0(\t,0,0,ia), \label{Kwl} \ee
is a holomorphic left mode which satisfies the equations of motion
and constraints.

To obtain the boundary conditions we have to consider an imbedding of this
string in ${\bf M^4}$. It is known that such an imbedding may not be
performed retaining the $N=2$ supersymmetry \cite{OogVaf,GibPer}.\fn{The
real spacetimes preserving the $N=2$ supersymmetry have signature
$(2,2)$ (Kleinian) or $(4,0)$ (Euclidean). } Therefore,
the  $N=2$ supersymmetry has to be broken, at least partially
\cite{BurSup}.

On the other hand, we have to use these left null planes also for the light
cones emanating  from the points of the complex world line $X_0(\t)$, and
note that not all the left null planes reach the real slice.

The null vectors $K^\m=x^\m - X_0^\m(\t _L),$ which belong to the complex
light cone, have to connect the complex world-line to the real points
$x^\m \in {\bf M^4}.$ This condition gives a restriction on the
positions of $\t _L$, which may be easily obtained from
the complex retarded-time equation. Let us write the light cone equation
in the form
\be 0=K_\m K^\m = [\vec x - \vec X_0(\t _L)]^2 - (t - \t _L)^2. \label{KK}
\ee
We assume here that  the gauge
$X_0^0(\t _L) = \t _L = t_L +i\sigma _L $ is chosen and that the Kerr
source is in the rest.

This equation may be split into the retarded-advanced time equations
\be  (t - \t _L) = \pm \tilde r \ ,\label{retadv} \ee where $ \tilde
r = \sqrt{[\vec x - \vec X_0(\t _L)]^2} $ is the complex radial
distance which takes the simple form $\tilde r=r +i a\cos \theta$ in
the Kerr oblate spheroidal coordinate system. Summarizing
these relations one obtains from the retarded time equation
(\ref{retadv}) \be t_L + i \sigma _L = t - r - ia \cos \theta
\label{tLcos}. \ee The real part of this equation, $t_L = t-r$, is
similar to the usual
retarded-time relation, while the imaginary part,
\be\sigma _L = -a\cos \theta \label{slth}, \ee
relates $\sigma _L$ with angle $\theta$.
Recall, that on the real slice the
projective coordinate $Y$ has the form
$Y=e^{i\phi} \tan \frac \theta 2$ and fixes the
angular directions of twistors.
Therefore, the parameter $\sigma _L$ fixes $\theta$ direction
leaving the coordinate $\phi$ free. One sees that there are the
limiting real rays corresponding to polar directions
$\theta =0 $ and $\theta =\pi$
which correspond to $\sigma _L= \pm a$. Only the light cones which
are joined to the interval $\sigma _L \in \Sigma _L=[-a,\ a]$ have
the real slice, and therefore, {\it only a part of the world-sheet,
the strip $\sigma \in [-a, \ a]$, is  `seen' at the real slice} and
may be imbedded in $\bf M^4$. It means that the world sheet
$(t,\sigma)$
 acquires the boundary, forming {\it an open complex string}.
The end point $\sigma_L = - a$ has $\theta =0$, and the
corresponding null plane $Y$ has the real slice along the positive
semiaxis $z^+$, while the end point $\sigma_L = + a$ has $\theta
=\pi$ and is mapped on the negative semiaxis $z^-$. Note, that these
semiaxis are exactly the singular semistrings of the real axial
stringy system.
Therefore {\it the complex string turns out to be a
D-string which is stuck to two singular semi-strings of opposite chiralities.}
Therefore, $z^\pm$ singular strings may carry the Chan-Paton factors
(chiral currents and traveling waves) which will play the role of quarks
with respect to the complex string.\fn{The relations of the $N=2$
string to the singular cosmic string was discussed for Kleinian spacetimes
in\cite{GibPer}. It was mentioned that Kleinian singularities can be
blown up, leading to the complete nonsingular metrics, by contrast in the
Minkowskian case the stringy singularities are topologically
nontrivial.}
\begin{figure}[ht]
\centerline{\epsfig{figure=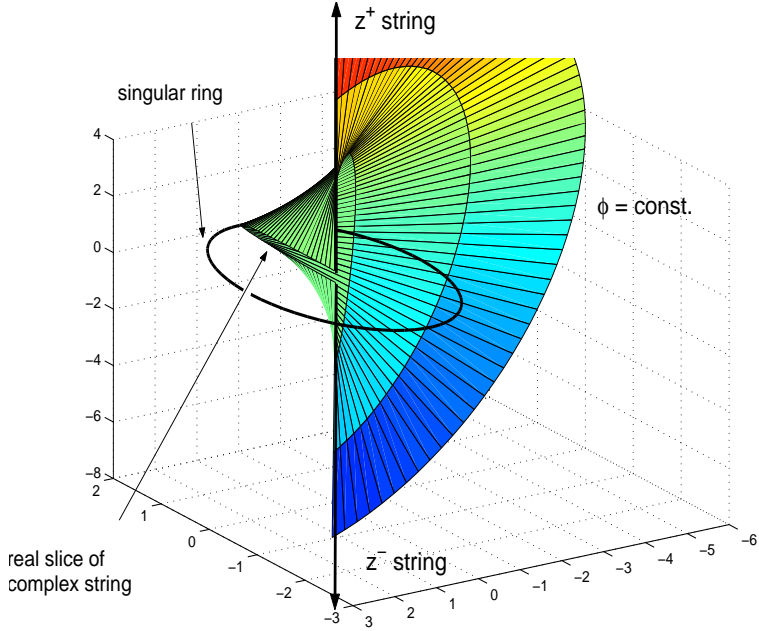,height=8.5cm,width=10cm}}
\caption{The complex twistor-string is imbedded into the real Kerr geometry.
The subset of semi-twistors ($r>0$) generating the Kerr angular coordinate
$\phi=const.$ is shown. The full set of semi-twistors may be obtained by the
rotation around z-axis}
\end{figure}

All the intermediate points of the interval $\sigma
\in [-a, \ a]$ have also the joined twistor planes $Y(\sigma_L)$ and
the joined real family of semitwistors on the real slice, see Fig. 4.

It should be noted that the $(x, \theta)$ string
was discussed in the papers \cite{Ber,LecPop,Nai,Sie2} as a
prospective alternative for the twistor-string.

\subsection{Orientifold}

The considered open twistor-string  has, however, a problem with
boundary conditions since they may not be defined for the real and
imaginary parts simultaneously.
 This problem
is resolved by orientifolding the string, which  involves the
complex conjugate structure - antiholomorphic complex world line
$\bar X_0^\m(\bar \t)$ and corresponding twistors $\{\bar
X_0^\m(\bar \t), \ \bar Y \}$. This structure is antichiral and
carries only the  right modes. The chiral and antichiral structures
may be joined by orientifolding the complex world-sheet
\cite{BurStr}.

Similar to the left interval $\sigma _L \in \Sigma _L=[-a,\ a]$, the
right structure has the right  interval $\sigma _R \in \Sigma
_R=[-a,\ a]$. However, in the right structure world-sheet has an opposite
orientation: the end point $\sigma _R = -a$ is mapped on the
negative semiaxis $z^-$, while the end point $\sigma _R = +a$ is
mapped on the positive semiaxis $z^+$.
Formation of the orientifold has a few steps.

The interval $\Sigma_L$, parametrized by $0>\theta
> \pi $, has to be extended to $0>\theta > 2\pi $.
The interval $\Sigma_R$ is reversed $\Sigma_R \to \bar
\Sigma_R$ and joined to the interval $\Sigma_L$. Therefore, the
new interval $\Sigma = \Sigma _L \cup \bar \Sigma _R $
covers twice the initiate
one going the second time in opposite direction and forming a closed loop.

The string modes are extended to $\pi >\theta > 2\pi $ in the standard manner
\cite{GSW} which allows one to consider the open string as a closed one:
\be X_L(\theta + \pi) = X_R(\theta), \quad  X_R(\theta + \pi) = X_L(\theta).
\ee
This operation is accompanied by the
transformations of the null planes $\bar Y\to -1/Y$ ($right \to left$)
and by the parity $r\to -r$ .

As a result the string acquires the right modes and
turns into a closed, but folded one  \cite{BurStr,BurOri,BurTwi}.
The final step is  the setting of the $\bf
Z_2$ equivalence of the left and right structures.

It should be noted that target space of this string turns out to be
equivalent to the `diagonal' of the $\bf CP^3 \times CP^{*3}$ which
was discussed by Witten in the end of paper \cite{WitTwi} and
also long ago \cite{WitYas} as a
formulation of the classical Yang-Mills theory.
The used in the Kerr spinning particle equivalent representation
$\{ X_0, \ Y \} $ may be preferable in some cases \cite{Sie2}, since the
spacetime coordinates appear explicitly in this form and geometrical
interpretation looks more transparent.

The extra complex coordinate $Y$ is free  for the complex $N=2$
string, but it acquires the $\sigma$ dependence by imbedding to the real
slice. In the $\bf CP^3$ representation of twistor-string it plays the role
of a target coordinate.
However, as it was discussed in \cite{OogVaf}, there are some
evidences that the world-sheet of the $N=2$ string is dual to its target
space, and is indeed four-dimensional.
It suggests that $Y$ may play a  double role: of the world-sheet and
target coordinate, which turns the $N=2$ string into a membrane as it was
discussed in \cite{OogVaf,Ket}. Similar, in it was suggested in the work
\cite{LecPop} that coordinate $Y$ must not
acquire the stringy excitations, and being to considered in the frame of
the string field theory, acts as a carrier of the field excitations.
These conjectures are supported by the Kerr spinning
particle.

\section{Conclusion}

A few related important problems have been left without treatment in
this paper. Among them are the supergeneralization of the model and
twist of the spin-structure.

A scheme for supergeneralization of the Kerr solution was considered
 in \cite{BurSup}, and it was shown  that the $N=2$ supersymmetry
 turns out to be broken in $\bf M^4$, which yields a residual nonlinear
realization of supersymmetry in the frame of the broken $N=2$ supergravity.
Therefore, the treatment of the supersymmetry in this model turns
out to be more complicated that in the Witten-Ferber approach
\cite{WitTwi,WitYas}.

 The twist of spin-structure \cite{WitTwi,Nai,BerVaf}
has to be incorporated in the model of the
Kerr spinning particle to provide a natural description of fermions
in terms of the half-integer excitations. In the Kerr
geometry twist is connected with topological structure of the axial
singular strings \cite{Egu}\fn{Note, that the axial
singularities of the Kerr spinning particle are different from
the black hole singularities considered in \cite{Egu}}, which turns out
to be related to the deep questions of the topological field theory
\cite{BerVaf,Egu,Wit3}. On the other hand, twist is close related to
supersymmetry. Preliminary treatment
shows that the Kerr-Schild formalism is {\it perfectly} matched to
twist, however, some basic relations have to be recalculated, which
demands a special  treatment.

Both these problems we intend to discuss elsewhere.

 \bigskip

 {\small Author is thankful to Organizing Committee of the
 Spin04-Praha Conference for invitation and financial support, and
 also to the participants of the Conference SPIN04-Praha and the
Trieste Symposium SPIN2004: A. Efremov, O. Teryaev, A. Dorokhov and
O. Selugin for interest in this work and very useful conversations.
Author would also like to thank  B. Dubrovin and A. Dabholkar  for
the useful discussions at Trieste.}

 \end{document}